# Large birefringence and linear dichroism in TiS$_3$ nanosheets

Nikos Papadopoulos[1*], Riccardo Frisenda[2], Robert Biele,[3] Eduardo Flores,[4] Jose R. Ares,[4] Carlos Sanchez,[4,5] Herre S. J. van der Zant,[1] Isabel J. Ferrer,[4,5] Roberto D'Agosta,[3,6*] and Andres Castellanos-Gomez[7*]

[1] *Kavli Institute of Nanoscience, Delft University of Technology, Lorentzweg 1, Delft 2628 CJ, The Netherlands.*
[2] *Instituto Madrileño de Estudios Avanzados en Nanociencia (IMDEA-Nanociencia), Campus de Cantoblanco, E-28049 Madrid, Spain.*
[3] *Nano-Bio Spectroscopy Group and European Theoretical Spectroscopy Facility (ETSF), Universidad del País Vasco UPV/EHU, 20018 San Sebastián, Spain.*
[4] *Materials of Interest in Renewable Energies Group (MIRE Group), Dpto. de Física de Materiales, Universidad Autónoma de Madrid, UAM, Campus de Cantoblanco, E-28049 Madrid, Spain.*
[5] *Instituto Nicolás Cabrera, Universidad Autónoma de Madrid, UAM, Campus de Cantoblanco E-28049 Madrid, Spain.*
[6] *IKERBASQUE, Basque Foundation for Science, 48013 Bilbao, Spain.*
[7] *Materials Science Factory, Instituto de Ciencia de Materiales de Madrid (ICMM-CSIC), Campus de Cantoblanco, E-28049 Madrid, Spain.*
*Email: n.papadopoulos@tudelft.nl, roberto.dagosta@ehu.es, andres.castellanos@csic.es

TiS$_3$ nanosheets have proven to be promising candidates for ultrathin optoelectronic devices due to their direct narrow band-gap and the strong light-matter interaction. In addition, the marked in-plane anisotropy of TiS$_3$ is appealing for the fabrication of polarization sensitive optoelectronic devices. Herein, we study the optical contrast of TiS$_3$ nanosheets of variable thickness on SiO$_2$/Si substrates, from which we obtain the complex refractive index in the visible spectrum. We find that TiS$_3$ exhibits very large birefringence, larger than that of well-known strong birefringent materials like TiO$_2$ or calcite, and linear dichroism. These findings are in qualitative agreement with *ab initio* calculations that suggest an excitonic origin for the birefringence and linear dichroism of the material.

**Introduction**

Since the isolation of graphene in 2004,[1] the list of 2D materials that can be exfoliated from bulk layered materials keeps growing rapidly.[2] Nowadays, anisotropic 2D materials, which are characterized by a strong in-plane anisotropy, are attracting the interests of the community. Moreover, materials whose optical properties strongly depend on the polarization of the incoming light are the foundation of many optical components (*e.g.* wave plates). In particular, polarimetric photodetectors are important for many applications such as ellipsometry, 3D imaging, non-destructive stress detection in industrial imaging and flat panel displays.[3] So far, the family of anisotropic 2D materials contains black phosphorous (BP),[4,5] Re-based chalcogenides,[6,7,8,9,10] tin sulfide[11] and selenide,[12] as well as transition metal trichalcogenides.[13,14,15,16,17] One of the most promising materials belonging to the latter class is $TiS_3$, which has a direct band-gap of 1.1 eV[18] and has shown very strong in-plane anisotropy[14,15,19] combined with exceptional responsivity and optoelectronic properties.[20,21] However, the optical anisotropy of $TiS_3$ has only been studied qualitatively and a the determination of the real and imaginary part of its refractive index is still lacking.[14]

In the present work, we investigate the optical properties of $TiS_3$ nanosheets in the visible part of the electromagnetic spectrum. In particular, we measure the optical contrast of $TiS_3$ nanosheets deposited on $SiO_2$/Si substrates as a function of the nanosheets thickness, ranging from 4 to 110 nm using unpolarized and linearly polarized light. The experimental results can be reproduced by a Fresnel law based model that takes into account the transmission and reflection phenomena taking place in the system (composed by a semi-infinite layer of air, $TiS_3$, 280 nm thick $SiO_2$ and semi-infinite Si). By fitting the experimental spectra, we extract the energy-resolved complex refraction index of $TiS_3$. We find both strong birefringence and linear dichroism, indicating that both the real and the imaginary part of the refractive index depend on the light polarization. Moreover, the results of *ab initio* calculations show that excitonic effects play a crucial role in the observed anisotropic optical properties of $TiS_3$. Interestingly, the birefringence coefficient of $TiS_3$ is among the largest values reported in literature[22] illustrating the potential of this novel 2D material for polarization optics applications.

**Results and discussion**

$TiS_3$ nanosheets are prepared by mechanical exfoliation of $TiS_3$ powder onto a polydimethylsiloxane (PDMS) film (Gelfilm from Gelpak®). The $TiS_3$ nanoribbons were synthesized by sulfuration of bulk Titanium discs. Titanium discs are vacuum sealed in an ampule with sulfur powder (>75 atomic % sulfur) and heated to a designated growth temperature (500 °C). After 20 hours of growth, the ampule is cooled in ambient conditions (see also Ref.[14]) and the as-produced $TiS_3$ powder can be transferred onto the PDMS film. The $TiS_3$ flakes are identified at first glance by optical inspection through a microscope operated in transmission mode and then deterministically transferred to the target 90 nm-$SiO_2$/Si substrate via a dry transfer technique.[23] Once deposited onto the $SiO_2$/Si substrates, the $TiS_3$ flakes are identified with bright-field optical microscopy and their thickness is determined by atomic force microscopy (AFM) measurements. Figure 1a shows an optical image of a thin $TiS_3$ nanosheet where two differently colored regions are visible. As can be seen from the nanosheet topography studied with AFM and shown in Fig. 1b, the regions with different colors have different thicknesses of approximately 4 and 20 nm, respectively. Once the $TiS_3$ nanosheets are deposited onto $SiO_2$/Si, their apparent color can range from red to blue, due to interference effects similar to other thin-film systems.[24,25] By studying several nanosheets with different thicknesses one can compile a color chart with the thickness-dependent apparent color of the $TiS_3$ (Fig. 1c-j). This color to thickness mapping can be used as a coarse approach to quickly estimate the thickness of $TiS_3$ through simple optical inspection without the need to perform further AFM measurements.[26,27] We refer the reader to Fig. S1 of the Supporting Information for optical and topographic images of additional thin $TiS_3$ nanosheets.

The optical contrast of $TiS_3$ flakes is quantitatively studied using a micro-reflectance setup described in detail elsewhere (see also Fig. S2 of the Supporting Information).[28] Briefly, the sample is illuminated by a white light source at normal incidence and the light reflected by the sample is collected with a fiber optic and fed to a CCD spectrometer. The area of the sample probed is a 1 μm diameter circular spot. To obtain the optical contrast of the $TiS_3$ nanosheet we perform two measurements, one in which we collect the light reflected from the $TiS_3$ nanosheet ($I_{fl}$) and a second for the light reflected by the $SiO_2$/Si substrate ($I_{sub}$). Figure 2a shows the wavelength-resolved optical contrast ($C$) of five flakes with different thicknesses, where the contrast is defined as $C=(I_{fl}-I_{sub})/(I_{fl}+I_{sub})$.[29] The contrast of a flake can take either positive (when the flake is brighter than the substrate) or negative values (flake darker than the substrate). In all the five cases displayed in Fig. 2a, the experimental contrast spectra present a modulation as a function of wavelength. This modulation is determined by interference conditions, which depend on the complex refractive index $\underline{n}=n+i\kappa$ and the thickness of the different stacked films ($TiS_3$ and $SiO_2$).

By performing tens of measurements, similar to the ones shown in Fig. 2a, on flakes with different thickness ranging from 4 nm to 110 nm, one can extract the optical contrast at a fixed wavelength as a function of the $TiS_3$ flakes thickness (see Fig. 2b-e). Note that the band structure of $TiS_3$ is expected to remain unchanged while varying the thickness.[30,31] These contrast *vs*. thickness datasets can be reproduced by a Fresnel law-based model[26] by using the two components of the complex refractive index of $TiS_3$ as free parameters and the refractive indexes of $SiO_2$ and Si from literature[32,33] as fixed parameters. The solid lines in figures 2b-e represent the theoretical optical contrast spectra which fit best the experimental data. By fitting contrast spectra for all the available wavelengths, we obtain the wavelength dependence of the complex refractive index of $TiS_3$.[29,34,35] Unlike other 2D materials, where quantum confinement effects have a strong influence on their band structure for flakes with different number of layers, the band structure of $TiS_3$ does not depend on the number of layers Ref. [30]. Therefore, it is expected that a single refractive index can describe well the properties of $TiS_3$ with different thicknesses.

Figure 3a-b shows the real and imaginary part of the refractive index ($n$ and $\kappa$) of $TiS_3$ extracted from the fitting procedure just described. The real part $n$ has values ranging from 3.2 to 4.2 for wavelengths between 475 and 700 nm, and its value increases for increasing wavelength. On the other hand, the imaginary part $\kappa$, which ranges between 0.5 and 1.1, increases as the wavelength decreases. The knowledge of the complex refractive index in van der Waals crystals allows the calculation of the optimal thickness of the $SiO_2$ layer on the Si for optical identification of monolayers. Figure 3c shows a colormap which represents the calculated optical contrast for a single layer $TiS_3$ (assuming a thickness of 0.9 nm according to the interlayer distance and in agreement with experimental observations[15]) as a function of the illumination wavelength (vertical axis) and $SiO_2$ thickness (horizontal axis). The optical contrast colormap shows maxima and minima and from these we can determine the $SiO_2$ thickness values that facilitate mostly the optical identification of a $TiS_3$ monolayer.[26,36,37] Considering that the human eye is more sensitive to light of 550 nm wavelength,[38] the optimal thicknesses of the $SiO_2$ layer would be approximately 80, 270 or 460 nm.

We now turn our attention to measurements employing linearly polarized light, which provide information about the in-plane anisotropy of $TiS_3$ nanosheets. Figure 4a shows a sketch of the $TiS_3$ monoclinic crystal structure taken along the *a*-axis direction. Due to structural anisotropy, $TiS_3$ grows preferentially along the *b*-axis, giving flakes which are elongated along that particular direction.[15] Figures 4b and 4c show bright-field optical images of $TiS_3$ nanosheets using linearly polarized light parallel to the *b*- and *a*-axis, respectively (as indicated in the figures). A difference in the color of the flakes can be seen when using two differently polarized light configurations. The optical contrast measured for different relative angle between the *b*-axis and the linearly polarized light is shown in Fig. 4d (and in Fig. S3 of the Supporting Information for additional flakes). As it can be seen, in the region between 530 and 570 nm the contrast of $TiS_3$ decreases as the polarization direction becomes parallel to the *a*-axis. The observed dependency of the contrast on the polarization of the incident light is the reason for the enhancement of the red color of the flake when the electric field is parallel to the *a*-axis. Moreover, we observed that flakes with a different thickness show a similar dependence of the contrast for polarization but centered at a different wavelength (see Fig. S3 of the Supporting Information).

By following the approach described above for unpolarized light, but now using linearly polarized light, the polarization dependent complex refractive index is extracted (see Fig. 5a-b). The real part of the refractive index is sensitive to the polarization angle of the incident light in the range 500 nm to 650 nm. The difference $\Delta n$ between $n_b$ and $n_a$, which are the refractive indices for linearly polarized light parallel to $b$ and $a$-axis, respectively, reaches $0.30 \pm 0.04$ at 560 nm. This value is much larger than that of other anisotropic van der Waals materials like black phosphorus, $ReS_2$ and $ReSe_2$ and it is also larger than that of well-known strongly birefringent materials like $TiO_2$,[22,39] calcite and barium borate.[39,40] Table 1 shows a comparison between the birefringence ($\Delta n$) of these materials. Also, Fig. 5b shows how the imaginary part of the refractive index is larger for light polarized along the $b$ direction and smaller for a polarization along the $a$ direction, indicating a stronger absorption along the $b$ direction. From these measurements we conclude that $TiS_3$ shows a marked linear dichroism along the whole visible spectrum. We attribute the small differences between the extracted unpolarized and polarized refractive indexes to the fact that in the case of the polarized light measurements a lower amount of $TiS_3$ flakes were studied. Nevertheless, the obtain polarized refractive indexes can reproduce quite well the experimental data (see Fig. S4). Figure 5c-d shows the polar plot of the real part of the refractive index at different illumination wavelengths for different polarization angles. The refractive index shows a lobe pattern with maxima and minima located at 90º and 0º respectively, which correspond to the $b$- and $a$-axis directions in the material.

To better understand our experimental findings, we performed state-of-the-art density functional theory (DFT) calculations in combination with many-body techniques. Considering the thickness of the sample, we have investigated the optical properties of bulk $TiS_3$. Further information about the atomic positions and other electronic properties can be found elsewhere.[41,42] Using DFT we first relaxed the geometry of the material, reaching a configuration where the residual forces between the atoms are small, and then calculated its electronic ground state. Afterwards, we performed $G_0W_0$ calculations in order to achieve a more accurate description of the electronic band structure and to go beyond some of the common know problems of DFT (small gap, lack of many-body effects etc.).[43] Finally, to study the optical properties we have calculated the optical spectrum within the random-phase approximation (RPA) and solved the Bethe-Salpeter equation (BSE). Both methods (RPA and BSE) include local field effects, accounting for macroscopic inhomogeneity of the charge density in the materials. While RPA takes only single-particle excitations into account, BSE includes electron-hole binding (excitonic) effects.

Figure 6 shows the real and the imaginary part of the refractive index of $TiS_3$ calculated within the RPA and the BSE methods. A direct comparison of the experimental results with the calculated optical properties shows qualitative agreement with the BSE calculations. Experimentally we found that the real part of the refractive index *n* is larger for polarization along the *a*-axis (Fig. 5a), while the imaginary part *κ* is smaller along the same axis (Fig. 5b). This behavior cannot be reproduced within the RPA approximation (see Fig. 6a), where excitonic effects are not considered. However, by including electron-hole binding effects (BSE, Fig. 6c and d) we can reproduce qualitatively both the strong birefringence and dichroism observed experimentally. This indicates that excitonic effects play a crucial role in determining the optical behavior of $TiS_3$, which is consistent with the large exciton binding energy of $TiS_3$.[18] However, in these calculations, a perfect quantitative agreement between theory and experiment cannot be expected in the observed energy range and the origin of this discrepancy can be manifold. For example, it could be a signature of lifetime effects due to electron-phonon interactions leading to a renormalization of the conduction bands.[41,42] Alternatively, it can be due to many-particle excitations involving more than two particles which are not described by the two-particle Hamiltonian we have solved. In this case bound triplet excitons (trions), which have been found in other 2D materials, might play a small but not negligible role in $TiS_3$ in the high-energy range probed by our experiment.

**Conclusions**

To summarize, we have studied the anisotropic optical properties of $TiS_3$ nanosheets in the visible spectrum. From wavelength-resolved micro-reflectance measurements of $TiS_3$ flakes with thickness ranging from 4 nm to 110 nm, the complex refractive index was extracted and used to calculate the thickness of $SiO_2$ that provides an enhanced visibility of the $TiS_3$ monolayer. Finally, we obtained the complex refractive index for light polarization parallel to the *a* and *b* crystallographic axis finding a remarkably strong birefringence, even larger than that of $TiO_2$ or calcite, well-known strongly birefringent materials.

**Conflicts of interest**

There are no conflicts to declare.

**Acknowledgements**

NP and HvdZ acknowledge support from the Organisation for Scientific Research (NWO) and the Ministry of Education, Culture, and Science (OCW) in the Netherlands. ACG acknowledges funding from EU Graphene Flagship (Grant Graphene Core2 785219)and from the European Research Council (ERC) under the European Union's Horizon 2020 research and innovation programme (grant agreement n° 755655, ERC-StG 2017 project 2D-TOPSENSE). RF acknowledges support from the Netherlands Organization for Scientific Research (NWO) through the research program Rubicon with project number 680-50-1515. RB and RDA acknowledge financial support by DYN-XC-TRANS (Grant No. FIS2013- 43130-P) and SElecT-DFT (Grant No. FIS2016-79464-P) of the Spanish Ministerio de Economia y Competitividad, the Grupo Consolidado UPV/EHU del Gobierno Vasco (IT578-13). RB acknowledges funding from the European Union's Horizon 2020 research and innovation programme under the Marie Skłodowska-Curie grant agreement No. 793318 MIRE Group thanks the financial support from MINECO-FEDER through the project MA2015-65203-R.

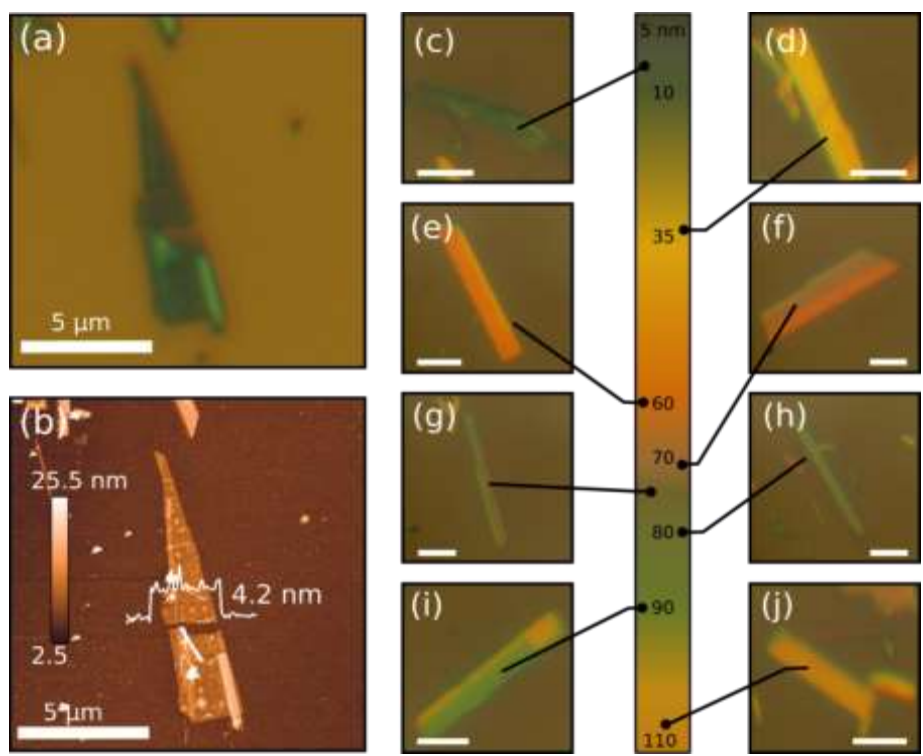

**Fig. 1**. Mapping the color of TiS$_3$ flakes to their thickness. (a) Optical image of an exfoliated thin flake of TiS$_3$ on 90 nm SiO$_2$/Si substrate. (b) AFM topography of the same flake from panel (a). The thickness of the particular flake is 4.2 nm. (c-j) Optical images of flakes with different thickness and a colorbar with the colors of flakes with thickness up to 110 nm. The scale bars are 5 μm.

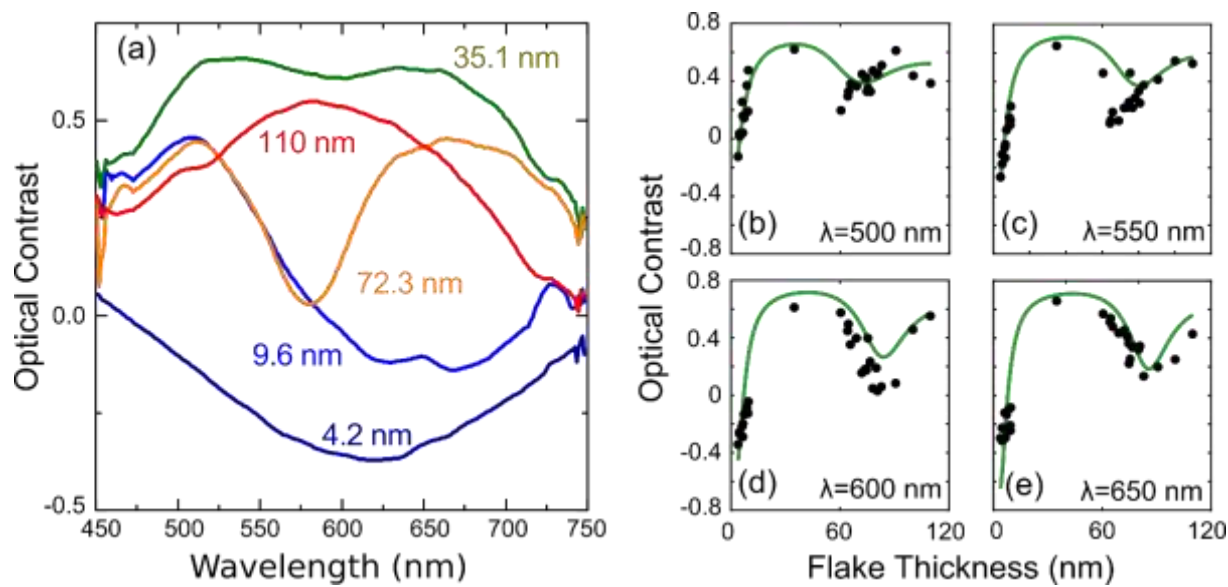

**Fig. 2**. Micro-reflectance measurements and determination of the complex refractive index. (a) Optical contrast as a function of wavelength for different TiS$_3$ flakes. (b-e) Optical contrast as a function of TiS$_3$ thickness at constant wavelength. Using the Fresnel model, the complex refractive index of TiS$_3$ can be determined. The green curves correspond to results of the fitting.

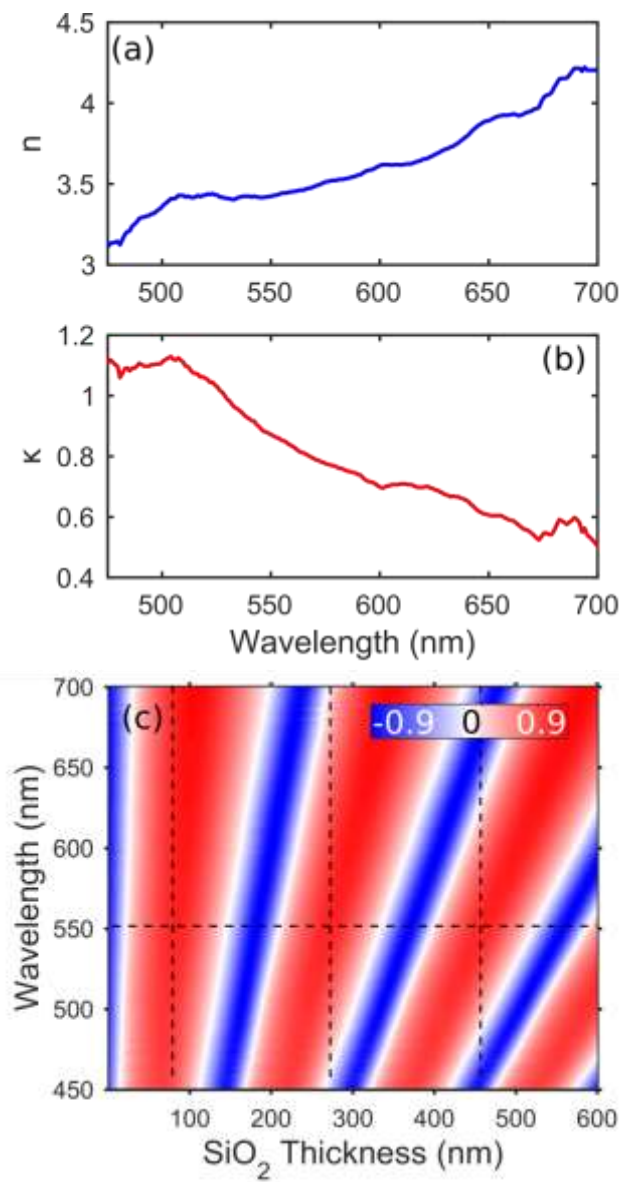

**Fig. 3**. Complex refractive index of TiS$_3$ and simulation of the optical contrast of a TiS$_3$ monolayer on SiO$_2$/Si. (a) Real and (b) imaginary part of the refractive index as a function of the wavelength. (c) Calculated optical contrast of a monolayer TiS$_3$ on SiO$_2$ for different wavelengths and SiO2 thicknesses, using the obtained values of the refractive index.

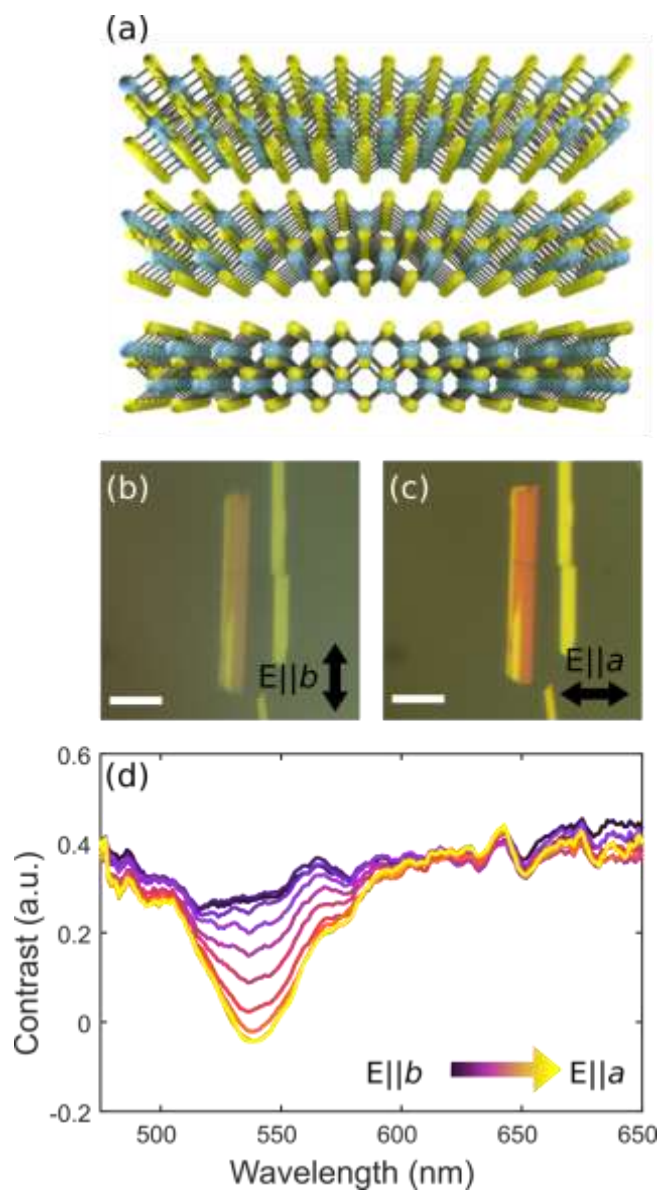

**Fig. 4**. Polarization dependent optical contrast of TiS$_3$. (a) Crystal structure of TiS$_3$. (b-c) Optical images of a 67.6 nm thick flake, obtained with the polarization axis of the light being parallel to the *b*-axis in (b) and *a*-axis in (c). The bars correspond to 7 µm. (d) Optical contrast spectra vs. wavelength at different polarization angles. The dark purple curve corresponds to light polarization parallel to the *b*-axis, while the yellow curve corresponds to polarization parallel to the *a*-axis.

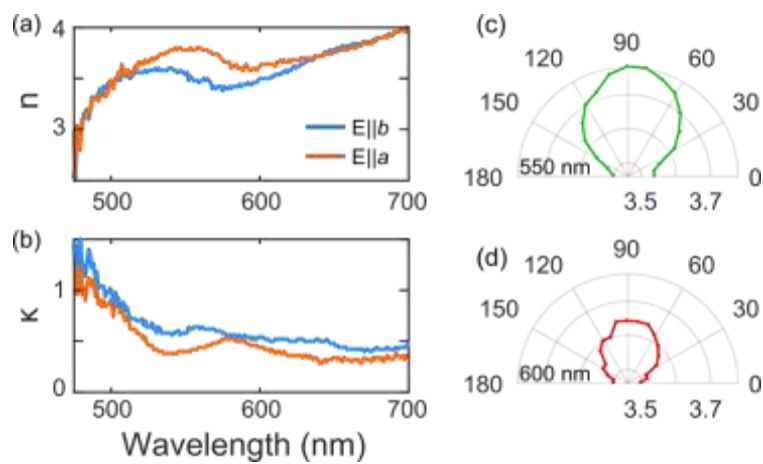

**Fig. 5**. Anisotropy of the refractive index in TiS$_3$. (a) Real part of the refractive index vs. wavelength for light polarization parallel to the *b* (blue curve) and *a* (red curve) axis. The maximum birefringence occurs for wavelengths around 560 nm. (b) Imaginary part as a function of wavelength for the two polarizations. The dichroism is taking place along the whole visible spectrum. (c-d) Polar plots of the real part of the refractive index for the wavelengths of 550 and 600 nm, respectively.

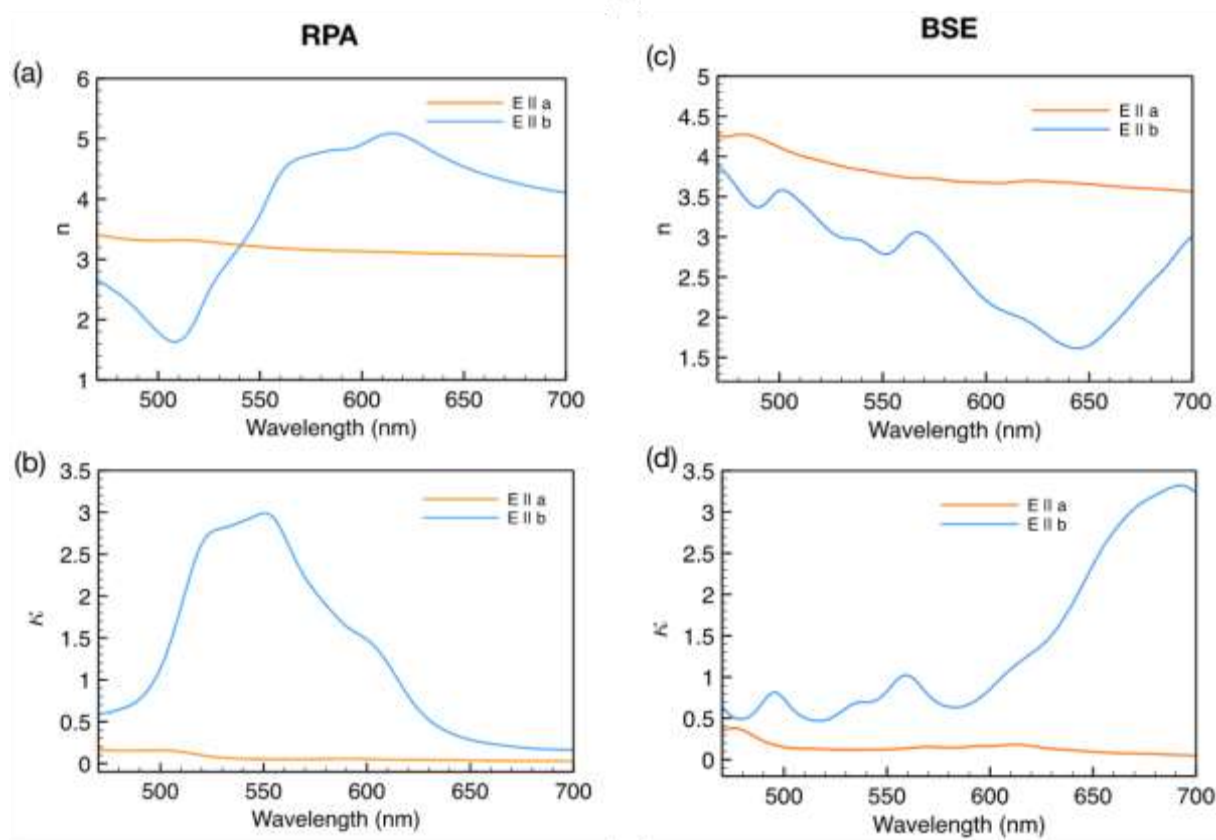

**Fig. 6** Refractive index obtained from *ab initio* calculations for polarizations parallel to *a*- and *b*-axis. In (a) and (b) the calculations are based on the random-phase approximation (RPA) that considers single particle excitations. In (c) and (d) the calculations are based on the Bethe-Salpeter equation (BSE) that includes excitonic effects. The convergence with the experiment takes place if the excitonic effects are included.

**Table 1.** Comparison between the birefringence magnitude of different birefringent materials and anisotropic van der Waals materials.

| Material | Δ$n$ | Reference |
|---|---|---|
| $TiS_3$ | +0.30 ± 0.04 | This work |
| $TiO_2$ rutile and hematite | +0.287 | 22,39 |
| Calcite $CaCO_3$ | -0.172 | 39 |
| Barium borate $BaB_2O_4$ | -0.1191 | 40 |
| Black phosphorus | +0.250 | 9 |
| ReS$_2$ | +0.037 | 9 |
| ReSe$_2$ | +0.047 | 9 |

**Supporting Information**

# Large birefringence and linear dichroism in TiS$_3$ nanosheets

Nikos Papadopoulos[1*], Riccardo Frisenda[2], Robert Biele,[3] Eduardo Flores,[4] Jose R. Ares,[4] Carlos Sanchez,[4,5] Herre S. J. van der Zant,[1] Isabel J. Ferrer,[4,5] Roberto D'Agosta,[3,6*] and Andres Castellanos-Gomez[7*]

[1] *Kavli Institute of Nanoscience, Delft University of Technology, Lorentzweg 1, Delft 2628 CJ, The Netherlands.*
[2] *Instituto Madrileño de Estudios Avanzados en Nanociencia (IMDEA-Nanociencia), Campus de Cantoblanco, E-28049 Madrid, Spain.*
[3] *Nano-Bio Spectroscopy Group and European Theoretical Spectroscopy Facility (ETSF), Universidad del País Vasco UPV/EHU, 20018 San Sebastián, Spain.*
[4] *Materials of Interest in Renewable Energies Group (MIRE Group), Dpto. de Física de Materiales, Universidad Autónoma de Madrid, UAM, Campus de Cantoblanco, E-28049 Madrid, Spain.*
[5] *Instituto Nicolás Cabrera, Universidad Autónoma de Madrid, UAM, Campus de Cantoblanco E-28049 Madrid, Spain.*
[6] *IKERBASQUE, Basque Foundation for Science, 48013 Bilbao, Spain.*
[7] *Materials Science Factory, Instituto de Ciencia de Materiales de Madrid (ICMM-CSIC), Campus de Cantoblanco, E-28049 Madrid, Spain.*
*Email: n.papadopoulos@tudelft.nl, roberto.dagosta@ehu.es, andres.castellanos@csic.es

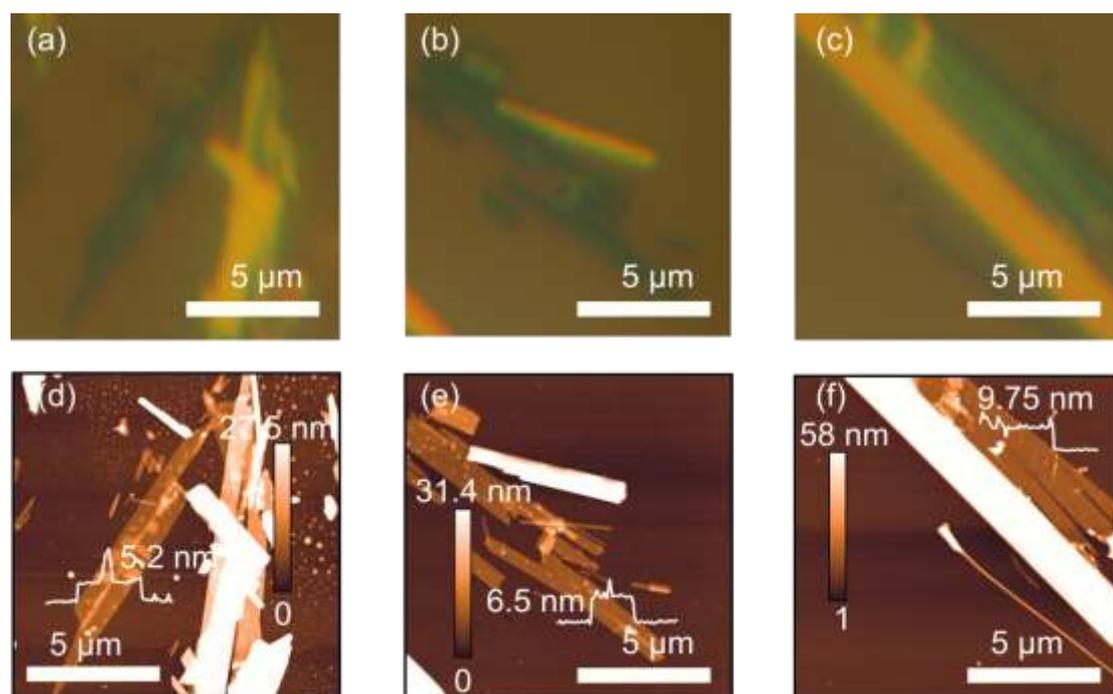

**Fig. S1**. Optical images (a, b and c) and AFM topography of the thin TiS$_3$ flakes (d, e and f). The thickness of these flakes is less than 10 nm.

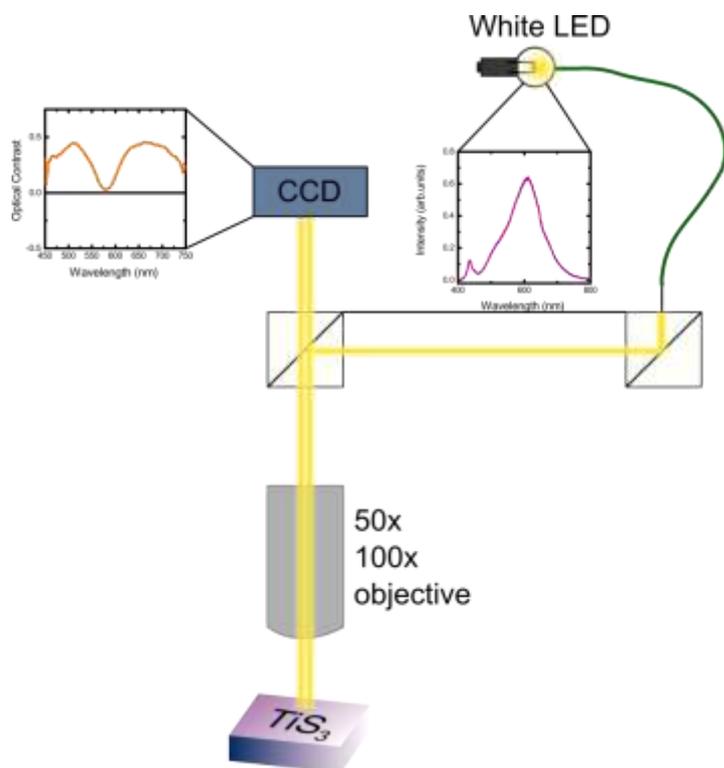

**Fig. S2**. Schematic of the Micro-reflectance setup used in this work.

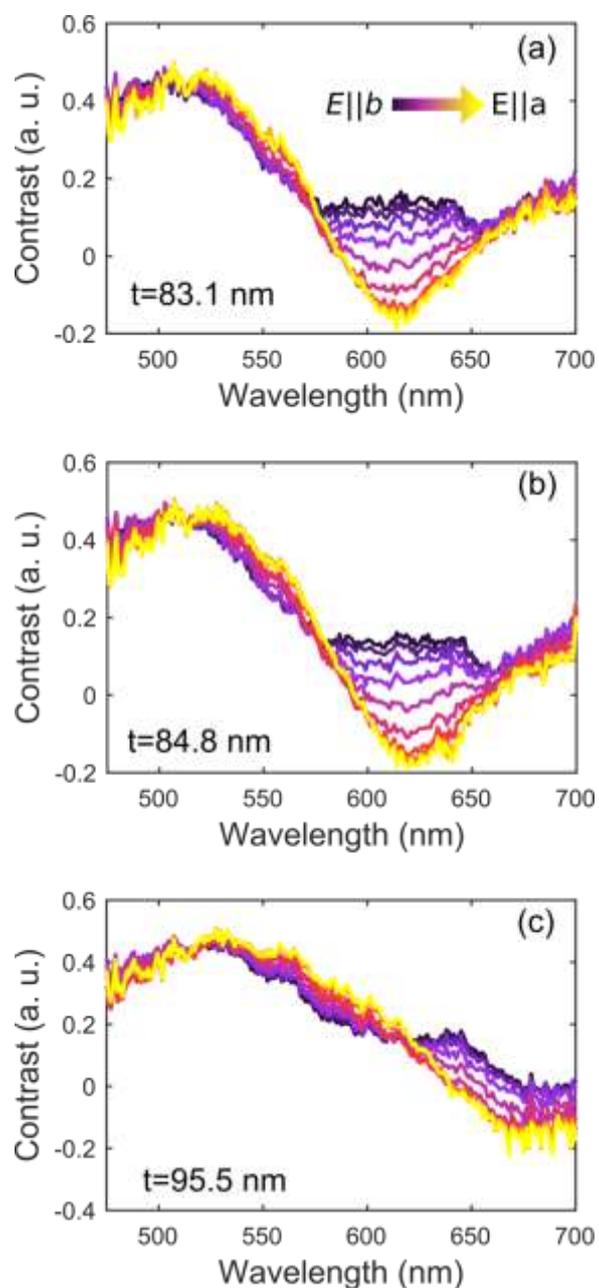

**Fig. S3**. (a-c) Contrast of various flakes as a function of wavelength for different polarization angles.

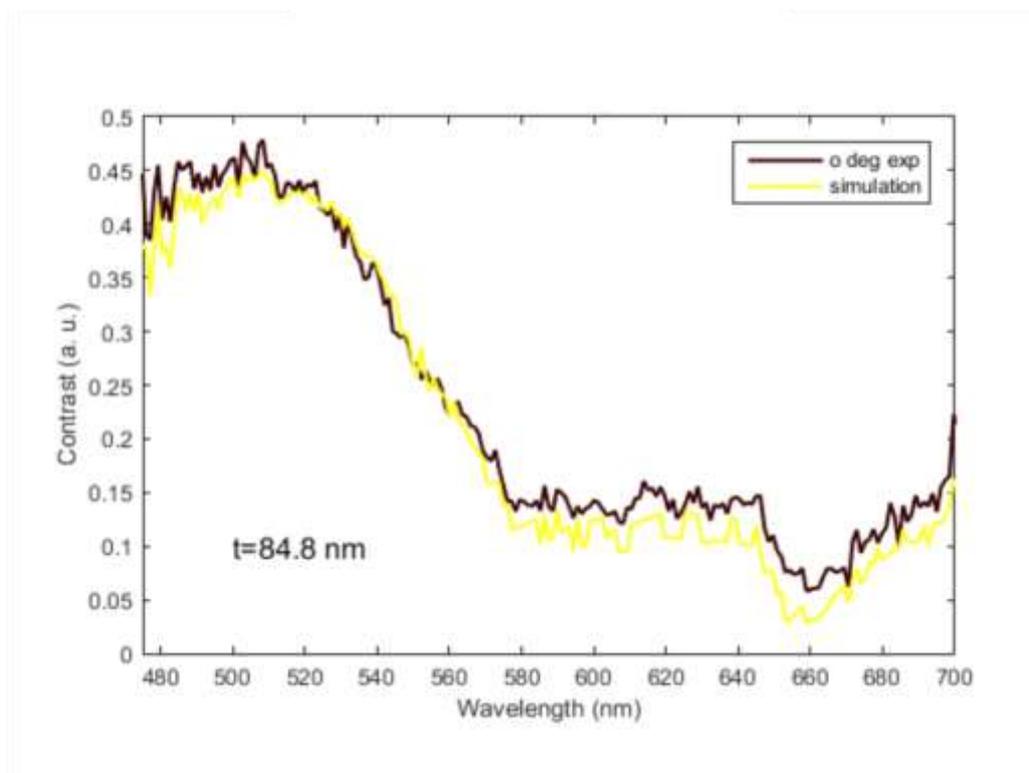

**Fig. S4**. Optical contrast spectra from a flake with thickness of 84.8 nm using linearly polarized light parralel to the *b*-axis. The yellow curve corresponds to the simulated spectra based on the Fresnel model using the obtained values of the refractive index that are shown in Fig.5.

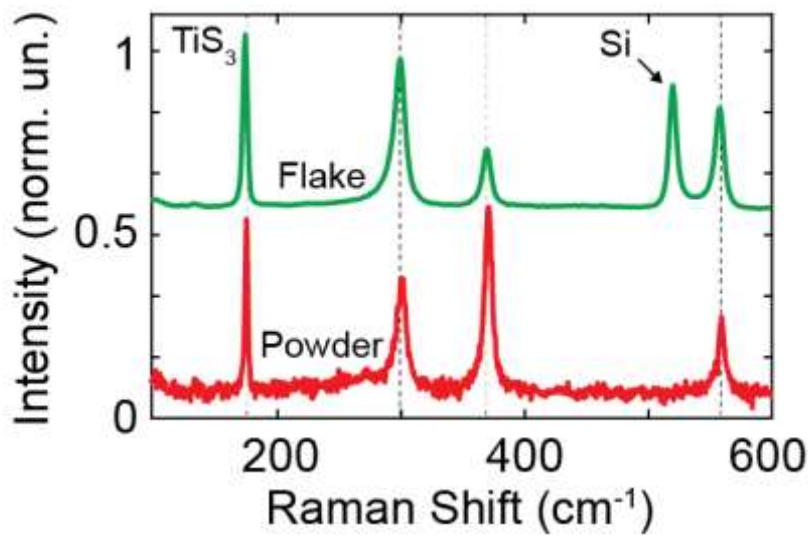

**Fig. S5**. Raman spectra of a TiS$_3$ nanoribbon on SiO$_2$/Si substrate (top) and of TiS$_3$ powder (bottom). The spectra show similar peaks due to TiS$_3$ vibrations. The spectrum of the individual TiS$_3$ flake shows an additional peak at 520 cm$^{-1}$ due to the Si substrate. While the energy of the peaks does not change when going from powder to an individual nanoribbon, the relative intensity of the peaks is different. This change is due to the different thicknesses probed in the powder sample compared to the individual flake.

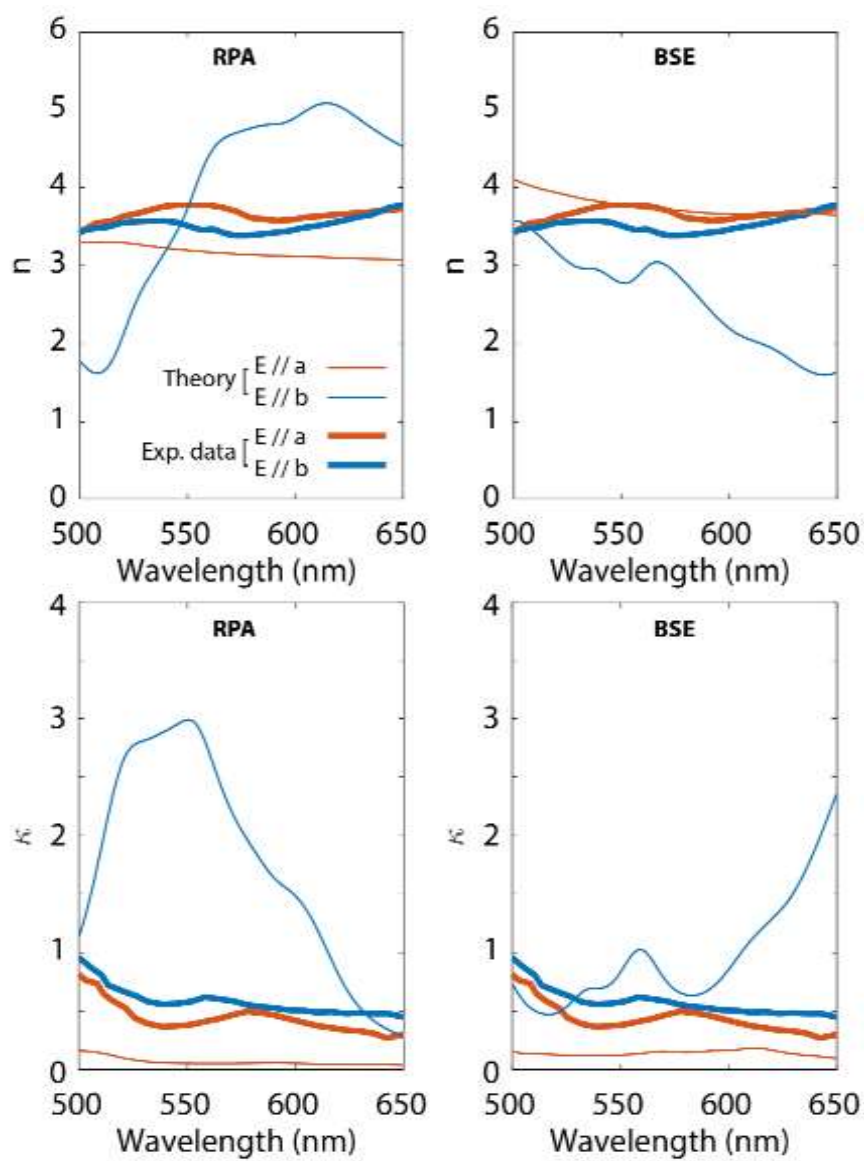

**Fig. S6.** Comparison between the two components of the polarized complex refractive index extracted from the measurements (thick lines) and calculated with *ab initio* calculations (RPA left, BSE right).